\documentclass[preprint,preprintnumbers,amsmath,amssymb]{revtex4}
\usepackage{amsmath}
\usepackage{txfonts}

\usepackage{graphicx}
\usepackage{rotating}
\usepackage{dcolumn}
\usepackage{bm}

\begin{document}


\title{Bosonization of Supersymmetric KdV equation}
\author{Xiao Nan Gao$^{1}$ and S Y Lou$^{1,2,3}$}
\affiliation{$^{1}$Department of Physics, Shanghai Jiao Tong
University, Shanghai, 200240, China\\
$^{2}$Faculty of Science, Ningbo University, Ningbo,
315211, China\\
$^3$School of Mathematics, Fudan University, Shanghai, 200433,
China}

\date{\today}

\begin{abstract}
Bosonization approach to the classical supersymmetric systems is presented. By
introducing the multi-fermionic parameters in the expansions of the
superfields, the $\mathcal {N}=1$ supersymmetric KdV (sKdV)
equations are transformed to a system of coupled bosonic equations. The method can be applied to
any fermionic systems.
By solving the coupled bosonic equations, some novel types of exact
solutions can be explicitly obtained. Especially, the richness of the localized excitations of the supersymmetric integrable system are discovered. The rich multi-soliton solutions obtained here have not yet been obtained by using other methods. Unfortunately, the traditional known multi-soliton solutions can also not be obtained by the bosonization approach of this paper. 
Some open problems on the bosonization of the supersymmetric integrable models are proposed in the both classical and quantum levels.
\end{abstract}


\maketitle

\subsection{Introduction}

It is known that the supersymmetric integrable systems are very important in many physical fields especially in quantum field theory and cosmology such as superstring theory  where it appears as a basic part of the string worldsheet physics or the theory of two-dimensional solvable lattice models, e.g., tricritical Ising models \cite{McA,Kulish}. Though the supersymmetric integrable systems have been studied by many authors in the both quantum and classical levels, various important
problems are still open. For instance,
in the usual quantum field theory, the bosonization approach is one of the
powerful methods which simplifies the procedure to treat complex
fermionic fields \cite{qf}. However, in our knowledge, there is no method to find a proper bosonization procedure for both quantum and classical supersymmetric
integrable models. To treat the
integrable systems with fermions such as the super integrable
systems \cite{super}, supersymmetric integrable systems \cite{SS}
and pure integrable fermionic systems \cite{fer} is much more
complicated than to study the integrable pure bosonic systems. Therefore, it is significant if one can establish a proper bosonization procedure to treat the supersymmetric systems even if in the classical level.

In this paper, taking the $N=1$ classical supersymmetric KdV (sKdV) system as a simple example we  propose a simple bosonization approach to find exact solutions of
supersymmetric systems. Actually, the method has been used by Andrea et al
\cite{0} to obtain new integrable bosonic systems. Here, we apply
the method to find new exact solutions of supersymmetric integrable
systems. One essential advantage of the
method is that it can effectively avoid difficulties caused by
intractable fermionic fields which are anticommuting. The $\mathcal
N =1$ supersymmetric versions of the Korteweg-de Vries equation have
been found more than $20$ years ago \cite{2,3,4}, which are the
beginning of the field of supersymmetric integrable systems. The
far-reaching significance lies not only in mathematics, but also in
the applications in various areas of modern theoretical physics. Therefore,
investigating their properties and searching for their exact
solutions are of great importance and interest.

For the integrable sKdV system in the sense of possessing a Lax
pair, many remarkable properties have been discovered, such as the
Painlev\'e property \cite{5}, the bi-Hamiltonian structures
\cite{6,7}, the Darboux transformation \cite{8}, the bilinear forms
\cite{9,10}, the B\"acklund transformation (BT) \cite{12}, the Lax
representation \cite{13} and the nonlocal conservation laws
\cite{14}. Some types of multisoliton solutions are also known for
the integrable sKdV system \cite{9,10,11,12,13}. However, because
anticommutative fermionic fields bring some difficulties in dealing
with supersymmetric equations, to get exact solutions of the
supersymmetric systems is, especially, much more difficult than the
usual pure bosonic systems.

The $\mathcal {N}=1$ supersymmetric version of the KdV equation,
\begin{eqnarray}\label{kdv}
u_t +6uu_x +u_{xxx} =0,
\end{eqnarray}
is established by extending the classical spacetime ($x, t$) to a
super-spacetime ($\theta, x, t$), where $\theta$ is a Grassmann
variable, and the field $u$ to a fermionic superfield
\begin{eqnarray}
\Phi(\theta, x, t) = \xi(x, t)+\theta u(x, t),
\end{eqnarray}
which leads to a nontrivial result \cite{3}
\begin{eqnarray}\label{Phi}
\Phi_t +3 (\mathcal {D} \Phi_x) \Phi +3 (\mathcal {D} \Phi) \Phi_x
+\Phi_{xxx} =0,
\end{eqnarray}
where $\mathcal {D} = \partial _{\theta} +\theta \partial _x$ is the
covariant derivative. The component version of Eq. \eqref{Phi} reads
\begin{subequations}\label{uxi}
\begin{equation}\label{u}
u_t +u_{xxx} -3\xi \xi_{xx} +6uu_x =0,
\end{equation}
\begin{equation}\label{xi}
\xi_t +\xi_{xxx} +3u_x \xi +3 u\xi_x =0,
\end{equation}
\end{subequations}
where $u$ and $\xi$ are bosonic and fermionic component fields,
respectively. Vanishing $\xi$ in Eq. \eqref{uxi}, only the usual
classical KdV equation remains.

Previous studies of the sKdV system were all directly based on Eq.
\eqref{Phi} or \eqref{uxi}. In this paper, we are only concentrated
on bosonization of the sKdV equations by expanding the supperfields
with respect to the multi-fermionic parameters. In the next section,
we present the bosonization approach of the sKdV system, in which
the superfields are expanded about two fermionic parameters. And
then the general traveling (in the usual space-time) periodic wave solutions, including the
solitary waves as special cases, of the model are found. some special
types of nontraveling (in the usual space-time) wave solutions (including all possible exact
solutions of the usual KdV equation) are also obtained. In sections
\ref{a03} and \ref{a04}, we extend the bosonization approach of the sKdV system
to the case of three fermionic parameters and $n$ fermionic parameters respectively. The last section is a
short summary and discussion.

\subsection{Two-fermionic-parameter bosonization} \label{a02}

Firstly, we expand the component fields $\xi$ and $u$ in the form of
\begin{subequations}\label{uxi2}
\begin{equation}\label{u2}
\xi(x, t) =p \zeta_1 +q \zeta_2 ,
\end{equation}
\begin{equation}\label{xi2}
u(x, t) =u_0 +u_1 \zeta_1 \zeta_2 ,
\end{equation}
\end{subequations}
where $\zeta_1$ and $\zeta_2$ are two Grassmann parameters, while
the coefficients $p\equiv p(x,\ t)$, $q\equiv q(x,\ t)$, $u_0\equiv
u_0(x,\ t)$ and $u_1\equiv u_1(x,\ t)$ are four usual real or
complex functions with respect to the spacetime variables $x$ and
$t$, then the sKdV system \eqref{u}--\eqref{xi} is changed to
\begin{subequations}\label{bos}
\begin{equation}\label{eu}
{u_0}_t +{u_0}_{xxx} +6u_0{u_0}_x =0,
\end{equation}
\begin{equation}\label{pt}
p_t +p_{xxx} +3 u_0p_x +3{u_0}_x p =0,
\end{equation}
\begin{equation}\label{qt}
q_t +q_{xxx} +3 u_0q_x +3{u_0}_x q =0,
\end{equation}
\begin{equation}\label{u1t}
{u_1}_t +{u_1}_{xxx} +6u_0{u_1}_x+6{u_0}_xu_1 =3(pq_{xx} -qp_{xx})
\end{equation}
\end{subequations}
that is just the bosonic-looking of the sKdV system \eqref{uxi} in
two fermionic parameter case. Eq.
\eqref{eu} is exactly the usual KdV equation which has been widely
studied. Eqs. \eqref{pt} and \eqref{qt} are linear homogeneous in
$p$ and $q$ respectively, and Eq. \eqref{u1t} is linear
nonhomogeneous in $u_1$. Thereby, in principle, these equations can
be easily solved. This is just one of the advantages of the
bosonization approach.

Now let us consider the traveling wave solutions of the
bosonic-looking equations \eqref{bos}. Introducing the traveling wave variable
$X =k x+ \omega t+c_0$ with constants $k$, $\omega$ and $c_0$, the system \eqref{eu}--\eqref{u1t} are
transformed to the ordinary differential equations (ODEs)
\begin{subequations}
\begin{equation}\label{travel_u0}
k^3 {u_0}_{XXX} +(6ku_0 +\omega) {u_0}_X =0,
\end{equation}
\begin{equation}\label{travel_p}
k^3 p_{XXX} +(3ku_0+\omega) p_X +3k{u_0}_Xp=0,
\end{equation}
\begin{equation}\label{travel_q}
k^3 q_{XXX} +(3ku_0+\omega) q_X +3k{u_0}_Xq=0,
\end{equation}
\begin{equation}\label{travel_u1}
k^3{u_1}_{XXX} +(6ku_0+\omega) {u_1}_X  +6k{u_0}_Xu_1 =3k^2(pq_{XX}
-qp_{XX}).
\end{equation}
\end{subequations}
\bf \em Remark. \rm The traveling waves in the superspace,
$
\Phi(x,\ t,\ \theta)=\Phi(kx+\omega t+c_0+\zeta\theta)
$, with Grassmann constant $\zeta$
are different from those of in the usual space-time $\{x,\ t\}$. Hereafter, the traveling waves we discuss are only in the usual space-time $\{x,\ t\}$
but not in the superspace $\{x,\ t,\ \theta\}$.

Obviously, Eq. \eqref{travel_u0} is the traveling wave reduction of
the KdV equation, and its periodic wave solutions including solitary
wave solutions are well known. To solve the ODE system
\eqref{travel_u0}--\eqref{travel_u1}, we try to build the mapping
and deformation relation between the traveling wave solutions of the
classical KdV equation and the sKdV equation, and then to construct
the exact solutions of the sKdV equation by using the known
solutions of the KdV equation.

We first solve out ${u_0}_X$ from Eq. \eqref{travel_u0}. The result
reads
\begin{eqnarray}\label{ru0}
{u_0}_X =\frac{a_0}{k^2} \sqrt{-k(2ku_0^3 +\omega
u_0^2+2c_1u_0-c_2k^3)},
\end{eqnarray}
where $c_1$ and $c_2$ are two integral constants and $a_0=\pm1$. The
only linear inhomogeneous ODE \eqref{travel_u1} can be directly
integrated once, and becomes
\begin{eqnarray}\label{tr_u1int}
k^3{u_1}_{XX} +(6ku_0+\omega)u_1 =f(X),
\end{eqnarray}
where the inhomogeneous term is
\begin{eqnarray}
f(X)=3k^2(pq_X-qp_X)-b_0
\end{eqnarray}
with an integral constant $b_0$.

To get the mapping relations of $p$, $q$ and $u_1$, we introduce the
variable transformations as follows
\begin{eqnarray}\label{trs_pqu1}
p(X)=P(u_0(X)), \qquad q(X)=Q(u_0(X)), \qquad u_1(X)=U_1(u_0(X)).
\end{eqnarray}
Using the transformation \eqref{trs_pqu1} and eliminating ${u_0}_X$
via Eq. \eqref{ru0}, the linear ODEs
\eqref{travel_p}--\eqref{travel_q} as well as \eqref{tr_u1int} are
changed to
\begin{subequations}\label{pqu0}
\begin{equation}\label{pu0}
(2ku_0^3+\omega u_0^2+2c_1u_0-c_2k^3)\frac{d^3P}{du_0^3}
+3(3ku_0^2+\omega u_0+c_1)\frac{d^2P}{du_0^2}
+3ku_0\frac{dP}{du_0}-3kP=0,
\end{equation}
\begin{equation}\label{qu0}
(2ku_0^3+\omega u_0^2+2c_1u_0-c_2k^3)\frac{d^3Q}{du_0^3}
+3(3ku_0^2+\omega u_0+c_1)\frac{d^2Q}{du_0^2}
+3ku_0\frac{dQ}{du_0}-3kQ=0,
\end{equation}
\begin{equation}\label{u1u0}
(2ku_0^3+\omega u_0^2+2c_1u_0-c_2k^3)\frac{d^2U_1}{du_0^2}
+(3ku_0^2+\omega u_0+c_1)\frac{dU_1}{du_0} -(6ku_0+\omega)U_1
=F(u_0),
\end{equation}
\end{subequations}
where
\begin{eqnarray}
F(u_0) =3a_0\left(Q\frac{dP}{du_0}-P\frac{dQ}{du_0}\right)
\sqrt{-k(2ku_0^3+\omega u_0^2+2c_1u_0-c_2k^3)} +b_0.
\end{eqnarray}
On this basis, the mapping and deformation relations are constructed
as
\begin{subequations}\label{rw}
\begin{eqnarray}
P(u_0) &=&A_1u_0+A_2\sqrt{ku_0^2+c_1}
\sin{\left[R(u_0)+A_3\right]},\\
Q(u_0) &=&A_4u_0+A_5\sqrt{ku_0^2+c_1} \sin{\left[R(u_0)+A_6\right]},
\end{eqnarray}
\end{subequations}
where $A_1$, $A_2$, $A_3$, $A_4$, $A_5$ and $A_6$ are arbitrary
constants, and $$ R(u_0)= \int{\frac{\sqrt{-c_1(c_2k^4+c_1\omega)}}
{(ku_0^2+c_1)\sqrt{2ku_0^3 +\omega u_0^2 +2c_1u_0-c_2k^3}}}\ du_0.
$$ Following the relation \eqref{rw}, the solution
for $U_1$ can be obtained from Eq. \eqref{u1u0} as
\begin{eqnarray}\label{ru1}
U_1(u_0) =\sqrt{2ku_0^3+\omega u_0^2+2c_1u_0-c_2k^3} \left[A_8
+\int{ \frac{A_7+\int{F(u_0)}du_0} {(2ku_0^3+\omega
u_0^2+2c_1u_0-c_2k^3)^{3/2}}} du_0\right] ,
\end{eqnarray}
where $A_7$ and $A_8$ are two integral constants. Thus, we have
obtained the general two-fermionic-parameter traveling wave
solutions of the sKdV system
\begin{subequations}\label{result1}
\begin{eqnarray}
u &=&u_0+\zeta_1\zeta_2 \sqrt{2ku_0^3+\omega u_0^2+2c_1u_0-c_2k^3}
\left[A_8 +\int{ \frac{A_7+\int{F(u_0)}du_0} {(2ku_0^3+\omega
u_0^2+2c_1u_0-c_2k^3)^{3/2}}} du_0\right]
,\\
\xi&=&\zeta_1\left\{A_1u_0+A_2\sqrt{ku_0^2+c_1}
\sin{\left[R(u_0)+A_3\right]}\right\}
+\zeta_2\left\{A_4u_0+A_5\sqrt{ku_0^2+c_1}
\sin{\left[R(u_0)+A_6\right]}\right\},\nonumber\\
&&
\end{eqnarray}
\end{subequations}
with the known solution $u_0$ of the usual KdV equation.

For a special case, $A_2=A_5=A_7=b_0=0$, the above traveling wave solution becomes
\begin{subequations}\label{Epq}
\begin{eqnarray}
P=a_1u_0, \ (a_1=A_1) \\
Q=a_2u_0, \ (a_2=A_4)
\end{eqnarray}
\end{subequations}
and
\begin{eqnarray}\label{Eu1}
U_1=A_8\sqrt{2ku_0^3+\omega u_0^2+2c_1u_0-c_2k^3} =a_3{u_0}_X,\
\left(a_3=\frac{A_8}{a_0}\sqrt{-k^3}\right),
\end{eqnarray}
where the second equal sign of the above equation is dues to the
relation \eqref{ru0}. It is interesting that the expression $U_1$
\eqref{Eu1} is an ordinary type of the symmetries of the traveling
wave equation \eqref{travel_u0}.

In fact, for any given $u_0(x,t)$ being a solution of the usual KdV equation, a certain type of solutions
of the bosonic-looking equation \eqref{bos} can be constructed as
follows
\begin{subequations}\label{epq}
\begin{eqnarray}
p&=&a_1u_0, \label{epqp} \\
q&=&a_2u_0, \label{epqq} \\
u_1&=&\sigma(u_0), \label{epqu1}
\end{eqnarray}
\end{subequations}
where $\sigma(u_0)$ represents any symmetry of the usual KdV
equation \eqref{eu}.

Under the circumstances of describing $p$ and $q$ as the form of
\eqref{epqp}--\eqref{epqq}, $u_0$ can be chosen as any solution of
the KdV equation. Then the first three equations of the
bosonic-looking equations \eqref{bos} are satisfied automatically.
Obviously, the righthand side of the nonhomogeneous equation
\eqref{u1t} equals zero because of Eqs. \eqref{epqp} and
\eqref{epqq}. In such situations, $u_1$ from Eq. \eqref{u1t} exactly
satisfies the symmetry equation of the usual KdV system \eqref{eu}.
This means that we have much freedom to choose $u_0$ so as to
construct solutions of the sKdV equations. Furthermore, it is not
limited to the traveling wave solutions of $u_0$. It is worth
mentioning that the KdV equation possesses infinitely many
symmetries, and thus infinitely many $u_1$ can be generated. All in
all, we can construct not only traveling wave solutions but also
many other new types of solutions of the sKdV system by using the
solutions and infinitely many symmetries of the KdV equation.

It is known that the solution \eqref{ru0} can be expressed by means
of the Jacobi elliptic functions, say,
\begin{eqnarray}
u_0=-\frac{k^3(2m^2-1)+\omega}{6k} +\frac{k^2m^2}{2}
\mbox{cn}^2\left(\frac{kx+\omega t+c_0}{2},m\right),
\end{eqnarray}
where the constants $c_1$ and $c_2$ are related to the other
constants through
$$c_1=\frac{\omega^2}{12k}-\frac{k^5}{12}(1-m^2+m^4),$$
and
$$c_2=-\frac{\omega^3}{108k^5}+\frac{k\omega}{36}(1-m^2+m^4)
-\frac{k^4}{108}(2-3m^2-3m^4+2m^6).$$

Therefore, we obtain a special type of exact solutions of the sKdV
system
\begin{subequations}\label{result1_1}
\begin{eqnarray}
u&=&-\frac{k^3(2m^2-1)+\omega}{6k} +\frac{k^2m^2}{2}
\mbox{cn}^2\left(\frac{kx+\omega t+c_0}{2},m\right)\nonumber\\
&&+\zeta_1\zeta_2\left\{C_1
J+C_2\big[1+3k^3m^2tJ\big]+C_3\big[2(\omega -k^3(m^2+1))+6k^3m^2S^2
\right.\nonumber\\
&&+3k^3m^2(3\omega t+kx)J\big]+C_4\big[18k^6m^4S^4-12k^3m^2(2\omega
+k^3(1+m^2))S^2
\nonumber\\
&&-2(m^2-2)(2m^2-1)k^6+8\omega
k^3(m^2+1)-4\omega^2+k^2\big(3k^2m^2(k^3(m^2+1)-4\omega)x
\nonumber\\
&& -3km^2(2k^6(1-m^2+m^4)+\omega k^3(2m^2-1)+8\omega^2)t-9k^5m^4\int
S^2 \mbox{\rm dx}\big)J\big]
\nonumber\\
&&\left.+C_5\big[6k^2m^3S^4-6mk^2S^2+(1-m^2)mk^2+3mk^2((E+m^2-1)(kx+\omega
t)+2Z)J\big]\right\},
\label{cn}\\
\xi&=&(a_1\zeta_1+a_2\zeta_2)\left[-\frac{k^3(2m^2-1)+\omega}{6k}
+\frac{k^2m^2}{2}\mbox{cn}^2\left(\frac{kx+\omega
t+c_0}{2},m\right)\right],
\end{eqnarray}
\end{subequations}
where $C_i\ (i=1,...,5),\ k,\ \omega,\ a_1,\ a_2,\ c_0,\ m$ are
usual arbitrary constants, $\zeta_1$ and $ \zeta_2$ are arbitrary
Grassmann odd constants, $Z\equiv Z\left(\frac12(kx+\omega
t+c_0),m\right)$ is the Jacobi zeta function, $E=\frac{E(m)}{K(m)}$
is the ratio of the complete elliptic integral of the second kind to
the first kind, and
$$S\equiv \mbox{sn}\left(\frac{kx+\omega t+c_0}{2},m\right),\qquad
J\equiv \mbox{cn}\left(\frac{kx+\omega t+c_0}{2},m\right)\mbox{dn}
\left(\frac{kx+\omega t+c_0}{2},m\right)S.$$ It is worth to mention
that the solution \eqref{cn} is neither a traveling nor a periodic
wave solution for $C_2C_3C_4\neq 0$.

It is noted that for the solution \eqref{result1_1}, the soliton
limit, $m=1, \ \omega=-k^3,$ exists for $C_2=C_5=0$. The result
reads ($\eta\equiv \frac{k}2(x-k^2 t-x_0),\ T\equiv \tanh(\eta)$)
\begin{subequations}\label{Sech}
\begin{eqnarray}
u&=&\zeta_1\zeta_2\left\{
\left[k(n_2+2n_3)x-k^3(3n_2+4n_3)t+n_1+n_3\ln\frac{1-T}
{1+T}\right]T-2n_2-2n_3\right\}
\mbox{sech}^2\left(\eta\right)\nonumber\\
&&+\frac{k^2}{2} \mbox{sech}^2\left(\eta\right),\label{sech}\\
\xi&=&(a_1\zeta_1+a_2\zeta_2) \mbox{sech}^2\left(\eta\right),
\end{eqnarray}
\end{subequations}
with the usually arbitrary constants $\{k,\ a_1,\ a_2,\ x_0,\ n_1,\
n_2,\ n_3\}$, and arbitrary Grassmann odd constants $\{\zeta_1,\
\zeta_2\}$.

The solution \eqref{uxi2} with \eqref{epq} extends every solution of
the KdV equation to many special types of solutions for the sKdV
system, due to the existence of the infinitely many symmetries of
the KdV equation. For instance, the N-soliton solution of the KdV
equation reads
\begin{equation}\label{NS}
u_{KdV}
=2\left\{\ln\left[1+\sum_{k=1}^N\sum_{i_1>i_2>\cdots>i_k}\prod_{m>n}A_{i_mi_n}
\exp\left(\sum_{j=1}^k\eta_{i_j}\right) \right]\right\}_{xx}
\end{equation}
with arbitrary constants $\{k_i,\ \eta_{0i},\ i=1,\ 2,\ ...,\ N\}$
and
$$A_{ij}=\frac{(k_i-k_j)^2}{(k_i+k_j)^2},\quad
\eta_j=k_jx-k_j^3t+\eta_{0j}.$$
Correspondingly, a special type of
multiple soliton solutions of the sKdV \eqref{uxi} can be simply
written as
\begin{subequations}\label{NSol}
\begin{eqnarray}
u&=&u_{KdV}+\zeta_1\zeta_2\sum_{i=0}^N\left(B_{i}
\frac{\partial}{\partial\eta_{0i}}u_{KdV}
+M_{i}\frac{\partial}{\partial k_{i}}
u_{KdV}\right),\label{nsol}\\
\xi&=&(a_1\zeta_1+a_2\zeta_2) u_{KdV}
\end{eqnarray}
\end{subequations}
with further arbitrary constants $B_i$ and $M_i$ for $i=1,\ 2,\
...,\ N$.

\subsection{Three-fermionic-parameter bosonization} \label{a03}

In the case of three Grassmann parameters $\zeta_1$, $\zeta_2$ and
$\zeta_3$, the component fields $\xi$ and $u$ are expanded as
\begin{subequations}\label{uxi3}
\begin{equation}\label{u3}
\xi(x, t) =p_1 \zeta_1 +p_2 \zeta_2 +p_3 \zeta_3 +p_4
\zeta_1\zeta_2\zeta_3 ,
\end{equation}
\begin{equation}\label{xi3}
u(x, t) =u_0 +u_1 \zeta_2 \zeta_3 +u_2 \zeta_3 \zeta_1 +u_3 \zeta_1
\zeta_2 ,
\end{equation}
\end{subequations}
where the coefficients $p_i\equiv p_i(x,\ t)$ $(i=1, 2, 3, 4)$ and
$u_j\equiv u_j(x,\ t)$  $(j=0, 1, 2, 3)$ are eight real or complex
bosonic functions of the indicated variables. Then the sKdV system
\eqref{u}--\eqref{xi} is changed to
\begin{subequations}\label{Eq}
\begin{equation}\label{Eq0}
{u_0}_t +{u_0}_{xxx} +6u_0{u_0}_x =0,
\end{equation}
\begin{equation}\label{Eq4}
{p_1}_t +{p_1}_{xxx} +3 u_0{p_1}_x +3{u_0}_x {p_1} =0,
\end{equation}
\begin{equation}\label{Eq5}
{p_2}_t +{p_2}_{xxx} +3 u_0{p_2}_x +3{u_0}_x {p_2} =0,
\end{equation}
\begin{equation}\label{Eq6}
{p_3}_t +{p_3}_{xxx} +3 u_0{p_3}_x +3{u_0}_x {p_3} =0,
\end{equation}
\begin{equation}\label{Eq1}
{u_1}_t +{u_1}_{xxx} +6u_0{u_1}_x+6{u_0}_xu_1 =3(p_2{p_3}_{xx}
-p_3{p_2}_{xx}),
\end{equation}
\begin{equation}\label{Eq2}
{u_2}_t +{u_2}_{xxx} +6u_0{u_2}_x+6{u_0}_xu_2 =3(p_3{p_1}_{xx}
-p_1{p_3}_{xx}),
\end{equation}
\begin{equation}\label{Eq3}
{u_3}_t +{u_3}_{xxx} +6u_0{u_3}_x+6{u_0}_xu_3 =3(p_1{p_2}_{xx}
-p_2{p_1}_{xx}),
\end{equation}
\begin{equation}\label{Eq7}
{p_4}_t +{p_4}_{xxx} +3 u_0{p_4}_x +3{u_0}_x {p_4} =-3 (u_1{p_1}
+u_2{p_2} +u_3{p_3} )_x .
\end{equation}
\end{subequations}
Just similar to the previous case, the system
\eqref{Eq0}--\eqref{Eq7} also has no fermionic quantities. Besides,
Eq. \eqref{Eq0} is exactly the KdV equation. The rest seven
equations are linear in $p_i$ $(i=1, 2, 3, 4)$, and $u_l$ $(l=1, 2,
3)$, respectively. It is observed that the number of the
inhomogeneous equations increases, so that this bosonic-looking of
the sKdV system is somewhat complex.

Introducing the traveling wave variable $X=kx+\omega t+c_0$, where
$k$, $\omega$ and $c_0$ are arbitrary constants, the bosonization
system \eqref{Eq} becomes
\begin{subequations}\label{trEq}
\begin{equation}\label{trEq0}
k^3{u_0}_{XXX} +(6ku_0+\omega) {u_0}_X =0,
\end{equation}
\begin{equation}\label{trEq4}
k^3{p_1}_{XXX} +(3ku_0+\omega) {p_1}_X +3k{u_0}_Xp_1=0,
\end{equation}
\begin{equation}\label{trEq5}
k^3{p_2}_{XXX} +(3ku_0+\omega) {p_2}_X +3k{u_0}_Xp_2=0,
\end{equation}
\begin{equation}\label{trEq6}
k^3{p_3}_{XXX} +(3ku_0+\omega) {p_3}_X +3k{u_0}_Xp_3=0,
\end{equation}
\begin{equation}\label{trEq1}
k^3{u_1}_{XXX} +(6ku_0+\omega) {u_1}_X  +6k{u_0}_Xu_1
=3k^2(p_2{p_3}_{XX} -p_3{p_2}_{XX}),
\end{equation}
\begin{equation}\label{trEq2}
k^3{u_2}_{XXX} +(6ku_0+\omega) {u_2}_X  +6k{u_0}_Xu_2
=3k^2(p_3{p_1}_{XX} -p_1{p_3}_{XX}),
\end{equation}
\begin{equation}\label{trEq3}
k^3{u_3}_{XXX} +(6ku_0+\omega) {u_3}_X  +6k{u_0}_Xu_3
=3k^2(p_1{p_2}_{XX} -p_2{p_1}_{XX}),
\end{equation}
\begin{equation}\label{trEq7}
k^3{p_4}_{XXX} +(3ku_0+\omega) {p_4}_X  +3k{u_0}_Xp_4 =
-3k(u_1{p_1}+u_2{p_2}+u_3{p_3})_X .
\end{equation}
\end{subequations}
It is quite obvious that Eq. \eqref{trEq0} is the same as
\eqref{travel_u0}, while Eqs. \eqref{trEq4}--\eqref{trEq6} have an
analogy with \eqref{travel_p}--\eqref{travel_q} and
\eqref{trEq1}--\eqref{trEq3} with \eqref{travel_u1}.  Coefficients
of the left-hand side of the last equation \eqref{trEq7} is
consistent with Eqs. \eqref{trEq4}--\eqref{trEq6}, but its
right-hand side is related to $p_l$ and $u_l$ $(l=1, 2, 3)$, not
always zero.

To solve the ODE system \eqref{trEq0}--\eqref{trEq7}, following the
approach adopted in the previous section, we first solve $p_l$ and
$u_l$. Integrating the inhomogeneous ODEs
\eqref{trEq1}--\eqref{trEq7} once, we have
\begin{subequations}\label{trEqint}
\begin{equation}
k^3{u_l}_{XX} +(6ku_0+\omega)u_l =f_l(X),
\end{equation}
\begin{equation}
k^3{p_4}_{XX} +(3ku_0+\omega)p_4=f_4(X),
\end{equation}
\end{subequations}
where
\begin{eqnarray}
f_1(X)&=&3k^2(p_2{p_3}_X-p_3{p_2}_X)-b_1 , \nonumber \\
f_2(X)&=&3k^2(p_3{p_1}_X-p_1{p_3}_X)-b_2 , \nonumber \\
f_3(X)&=&3k^2(p_1{p_2}_X-p_2{p_1}_X)-b_3 , \nonumber \\
f_4(X)&=&-3k(u_1p_1 +u_2p_2 +u_3p_3)-b_4 ,
\end{eqnarray}
with constants $b_1$, $b_2$, $b_3$ and $b_4$.

Considering the variable transformations
\begin{eqnarray}
u_l(X)&=&U_l(u_0(X)),\ (l=1, 2, 3), \nonumber \\
p_i(X)&=&P_i(u_0(X)), \ (i=1, 2, 3, 4)
\end{eqnarray}
and using \eqref{ru0} to eliminate ${u_0}_X$, we can transform the
linear ODEs \eqref{trEq4}--\eqref{trEq6} and \eqref{trEqint} to
\begin{subequations}\label{puu0}
\begin{equation}\label{plu0}
(2ku_0^3+\omega u_0^2+2c_1u_0-c_2k^3)\frac{d^3P_l}{du_0^3}
+3(3ku_0^2+\omega u_0+c_1)\frac{d^2P_l}{du_0^2}
+3ku_0\frac{dP_l}{du_0}-3kP_l=0,
\end{equation}
\begin{equation}\label{ulu0}
(2ku_0^3+\omega u_0^2+2c_1u_0-c_2k^3)\frac{d^2U_l}{du_0^2}
+(3ku_0^2+\omega u_0+c_1)\frac{dU_l}{du_0} -(6ku_0+\omega)U_l
=F_l(u_0),
\end{equation}
\begin{equation}\label{p4u0}
(2ku_0^3+\omega u_0^2+2c_1u_0-c_2k^3)\frac{d^2P_4}{du_0^2}
+(3ku_0^2+\omega u_0+c_1)\frac{dP_4}{du_0} -(3ku_0+\omega)P_4
=F_4(u_0) ,
\end{equation}
\end{subequations}
where
\begin{eqnarray}
F_1(u_0) &=&3a_0\left(P_3\frac{dP_2}{du_0}-P_2\frac{dP_3}{du_0}
\right)
\sqrt{-k(2ku_0^3+\omega u_0^2+2c_1u_0-c_2k^3)} +b_1\nonumber \\
F_2(u_0) &=&3a_0\left(P_1\frac{dP_3}{du_0}-P_3\frac{dP_1}{du_0}
\right)
\sqrt{-k(2ku_0^3+\omega u_0^2+2c_1u_0-c_2k^3)} +b_2, \nonumber \\
F_3(u_0) &=&3a_0\left(P_2\frac{dP_1}{du_0}-P_1\frac{dP_2}{du_0}
\right)
\sqrt{-k(2ku_0^3+\omega u_0^2+2c_1u_0-c_2k^3)} +b_3, \nonumber \\
F_4(u_0) &=&3k(U_1P_1 +U_2P_2 +U_3P_3 )+b_4.
\end{eqnarray}

By repeating the processes in the last section, the general
three-fermionic-parameter traveling wave solutions for the sKdV
system can be derived
\begin{eqnarray}\label{rpl}
u &=&u_0+u_{0X}\sum_{l=1}^3\zeta_l\zeta_{l+1}\left[h_{l} +\int{
\frac{g_{l}+\int{F_l(u_0)}du_0} {(2ku_0^3+\omega
u_0^2+2c_1u_0-c_2k^3)^{3/2}}}\ du_0\right] ,\nonumber \\
\xi &=&\sum_{l=1}^3\zeta_l\left\{r_{l}u_0+s_{l}\sqrt{ku_0^2+c_1}
\sin{\left[R(u_0)+\alpha_{l}\right]}\right\}\ \nonumber \\
&&+\zeta_1\zeta_2\zeta_3 \left\{r_{4}u_0+s_{4}\sqrt{ku_0^2+c_1}
\sin{\left[R(u_0)+\alpha_{4}\right]}\right.
\nonumber \\
&&\left.+ \sqrt{ ku_0^2 +c_1}
\int^{u_0}{\frac{\sin{\left[R(u_0)-R(y)\right]}F_4(y)\sqrt{ky^2+c_1}}
{\sqrt{c_1(c_2k^4+c_1\omega) (c_2k^3-2ky^3-\omega y^2-2c_1y)}}}\
dy\right\},
\end{eqnarray}
where $\{g_l,\ h_l,\ r_l,\ r_4,\ s_l,\ s_4,\ \alpha_l,\ \alpha_4\}$
are arbitrary constants and $\zeta_4=\zeta_1$.

Similar to the two fermionic parameter case, for nontraveling wave solutions of
Eq. \eqref{Eq}, we just write down a special case with
\begin{subequations}\label{rp2}
\begin{eqnarray}
p_l&=&d_lu_0,\ (l=1,\ 2,\ 3,\ 4),\\
u_1&=&\sigma_1(u_0),\\
u_2&=&\sigma_2(u_0),\\
u_3&=&-d_3^{-1}(d_1u_1+d_2u_2),
\end{eqnarray}
\end{subequations}
where $d_1,\ d_2,\ d_3,\ d_4$ are constants, $u_0$ is an arbitrary
solution of the usual KdV equation, $\sigma_1(u_0)$ and
$\sigma_2(u_0)$ are arbitrary symmetries of the usual KdV equation.
Finally, the sKdV system \eqref{uxi} possesses the following special
solution
\begin{subequations}\label{NSOL2}
\begin{eqnarray}
u&=&u_0+\sigma_1(u_0) \zeta_2 \zeta_3 +\sigma_2(u_0) \zeta_3 \zeta_1
-d_3^{-1}\big[d_1\sigma_1(u_0)+d_2\sigma_2(u_0)\big]
 \zeta_1 \zeta_2 ,\\\
\xi&=&(d_1\zeta_1+d_2\zeta_2+d_3\zeta_3+d_4\zeta_1\zeta_2\zeta_3)u_0
\end{eqnarray}
\end{subequations}
with an arbitrary solution $u_0$, two arbitrary symmetries
$\sigma_1(u_0)$ and $\sigma_2(u_0)$ of the usual KdV equation, three
Grassmann numbers $\zeta_i\ (i=1,\ 2,\ 3)$ and four arbitrary usual
real constants $d_l\ (l=1,\ 2,\ 3,\ 4)$. When one of the $\zeta_i$
tends to zero, the solution \eqref{NSOL2} turns back to that of the last
section for two fermionic parameters.

Actually, applying the similar procedure for any numbers of the
fermionic parameters, one can obtain various exact solutions such as
the general traveling wave solution and the special solutions like
Eq. \eqref{NSOL2}.

\subsection{N-fermionic-parameter bosonization} \label{a04}

For the supersymmetric system introduced $N\geq2$ fermionic
parameters $\zeta_{i}\ (i=1,2, \cdots, N)$, the component fields $u$
and $\xi$ can be expanded as
\begin{subequations}\label{uxiN}
\begin{equation}
u(x, t) =u_0 + \sum_{n=1}^{[\frac{N+1}{2}]} \sum_{1\leq i_{1}<
\cdots <i_{2n}\leq N} u_{i_{1}i_{2} \cdots i_{2n}} \zeta_{i_{1}}
\zeta_{i_{2}} \cdots \zeta_{i_{2n}} ,
\end{equation}
\begin{equation}
\xi(x, t) = \sum_{k=1}^{[\frac{N+1}{2}]} \sum_{1\leq i_{1}< \cdots
<i_{2n-1}\leq N} v_{i_{1}i_{2} \cdots i_{2n-1}} \zeta_{i_{1}}
\zeta_{i_{2}} \cdots \zeta_{i_{2n-1}},
\end{equation}
\end{subequations}
where the coefficients $u_{0}\equiv u_{0}(x,t)$, $u_{i_{1}i_{2}
\cdots i_{2n}}\equiv u_{i_{1}i_{2} \cdots i_{2n}}(x, t)\ (1 \leq
i_{1} <i_{2} < \cdots <i_{2n} \leq N)$ and $v_{i_{1}i_{2} \cdots
i_{2n-1}} \equiv v_{i_{1} i_{2} \cdots i_{2n-1}}(x, t)\ (1 \leq
i_{1} <i_{2} < \cdots <i_{2n-1} \leq N)$ are $2^N$ real or complex
bosonic functions of classical spacetime variable $x$ and $t$.
Substituting Eq. \eqref{uxiN} into the sKdV model \eqref{uxi}, we
obtain the following bosonic system of $2^N$ equations
\begin{subequations}\label{Eqn}
\begin{eqnarray}\label{Eqn-kdv}
{u_0}_t +{u_0}_{xxx} +6u_0{u_0}_x =0,
\end{eqnarray}
\begin{eqnarray}
L_o v_{i_1 i_2 \cdots i_{2n-1}} &=& \left\{\begin{array}{ll} 0
& \textrm{for $n=1$}\\
-3\sum\limits_{W_1} (-1)^{\tau(j_1,j_2, \cdots ,j_{2n-1})}
[u_{i_{j_1}i_{j_2} \cdots i_{j_{2l}}} v_{i_{j_{2l+1}}i_{j_{2l+2}}
\cdots i_{j_{2n-1}}}]_x & \textrm{for $n=2, 3, \cdots,
[\frac{N+1}{2}]$}
\end{array} \right. ,\label{Eqn-o}\\
L_e u_{i_1 i_2 \cdots i_{2n}} &=& \left\{\begin{array}{ll}
3\sum\limits_{W_2}(-1)^{\tau(j_1,j_2)} [v_{i_{j_{1}}}
(v_{i_{j_{2}}})_x]_x,
&\textrm{for $n =1$} \\
3\sum\limits_{W_2}(-1)^{\tau(j_1,j_2, \cdots ,j_{2n})}
[v_{i_{j_{1}}i_{j_{2}} \cdots i_{j_{2l-1}}}
(v_{i_{j_{2l}}i_{j_{2m+1}} \cdots i_{j_{2n}}})_x]_x \\
-3\sum\limits_{W_3} (-1)^{\tau(j_1,j_2, \cdots ,j_{2n})}
[u_{i_{j_1}i_{j_2} \cdots i_{j_{2l}}} u_{i_{j_{2l+1}}i_{j_{2l+2}}
\cdots i_{j_{2n}}}]_x     &\textrm{for $n=2, 3, \cdots,
[\frac{N}{2}]$}, \end{array} \right., \label{Eqn-e}
\end{eqnarray}
\end{subequations}
where
\begin{displaymath}
\tau(j_1,j_2, \cdots ,j_{N}) =\left\{\begin{array}{ll} 0 &
\textrm{$j_1,\ j_2,\ \cdots,\ j_{N}$ is even permutation}
\\ 1 & \textrm{$j_1,\ j_2,\ \cdots,\ j_{N}$ is odd permutation}
\end{array}\right. ,
\end{displaymath}
\begin{eqnarray}
W_1 &=&\{(j_1,j_2, \ldots j_{2n-1}) |1\leq j_1<j_2< \cdots < j_{2l}
\leq 2n-1, 1\leq j_{2l+1}<j_{2l+2}<\cdots < j_{2n-1} \leq 2n-1, \nonumber\\
&&1\leq l\leq n-1, j_{h_1}\neq j_{h_2}(h_1\neq h_2)\}, \nonumber\\
W_2 &=& \{(j_1,j_2, \ldots j_{2n}) |1\leq j_1<j_2< \cdots < j_{2l-1}
\leq 2n, 1\leq j_{2l}<j_{2l+1}<\cdots < j_{2n}\leq 2n, \nonumber\\
&&1\leq l\leq n, j_{h_1}\neq j_{h_2}(h_1\neq h_2)\}, \nonumber\\
W_3 &=& \{(j_1,j_2, \ldots j_{2n}) |1\leq j_1<j_2< \cdots < j_{2l}
\leq 2n, 1\leq j_{2l+1}<j_{2l+2}<\cdots < j_{2n}\leq 2n, \nonumber\\
&&1\leq l\leq n-1, j_{h_1}\neq j_{h_2}(h_1\neq h_2)\}, \nonumber
\end{eqnarray}
and two operators read
\begin{eqnarray}
L_e(u_0) &=&\partial_{t} +\partial_{xxx} +6u_0\partial_{x}
+6{u_0}_{x},\nonumber\\
L_o(u_0) &=&\partial_{t} +\partial_{xxx} +3u_0\partial_{x} +3{u_0}_{x}.
\nonumber
\end{eqnarray}

Similarly to the section, the general traveling wave solution of the
sKdV equation \eqref{Phi} with $N$ fermionic parameters can be
written as
\begin{equation}
\Phi =\sum\limits_{n=1}^{[\frac{N+1}{2}]} \sum\limits_{1\leq i_{1}<
\cdots <i_{2n-1}\leq N}v_{i_{1}i_{2} \cdots
i_{2n-1}}\zeta_{i_{1}}\zeta_{i_{2}} \cdots \zeta_{i_{2n-1}}
+\theta(u_0+\sum\limits_{n=1}^{[\frac{N+1}{2}]} \sum\limits_{1\leq
i_{1}< \cdots <i_{2n}\leq N}u_{i_{1}i_{2} \cdots
i_{2n}}\zeta_{i_{1}}\zeta_{i_{2}} \cdots \zeta_{i_{2n}}),
\end{equation}
where
\begin{subequations}
\begin{eqnarray}
v_{i_{1} i_{2} \cdots i_{2n-1}} &=& V_{i_{1} i_{2} \cdots
i_{2n-1}} (u_{0}) \nonumber\\
&=& r_{i_{1} i_{2} \cdots i_{2n-1}} u_0 +s_{i_{1} i_{2} \cdots
i_{2n-1}} \sqrt{ku_0^2+c_1} \sin[R(u_0)
+\alpha_{i_{1} i_{2} \cdots i_{2n-1}}] \nonumber\\
&&+\sqrt{ku_0^2+c_1} \int^{u_0} \frac{\sin[R(u_0)-R(y)] E_{i_{1}
i_{2} \cdots i_{2n-1}} (y) \sqrt{ky^2+c_1}}
{\sqrt{c_1(c_2k^4+c_1\omega) (c_2k^3-2ky^3-\omega y^2-2c_1y)}} dy,\\
u_{i_{1}i_{2}\cdots i_{2n}} &=& U_{i_{1} i_{2} \cdots i_{2n}} (u_{0})\nonumber\\
&=&\sqrt{2ku_0^3+\omega u_0^2+2c_1u_0-c_2k^3} \left[h_{i_{1}i_{2}
\cdots i_{2n}} +\int \frac{g_{i_{1}i_{2} \cdots i_{2n}} +\int
F_{i_{1}i_{2} \cdots i_{2n}}(u_0)du_0} {(2ku_0^3+\omega
u_0^2+2c_1u_0-c_2k^3)^{3/2}}du_0\right],\nonumber\\
&&
\end{eqnarray}
\end{subequations}
with
\begin{eqnarray}
E_{i_{1}i_{2} \cdots i_{2n-1}}(u_{0}) &=& \left\{
\begin{array}{ll}
0 & \textrm{if $n=1$}\\
3k\sum\limits_{W_1}(-1)^{\tau(j_1,j_2, \cdots, j_{2n-1})}
U_{i_{j_1}i_{j_2} \cdots i_{j_{2l}}} V_{i_{j_{2l+1}}i_{j_{2l+2}}
\cdots i_{j_{2n-1}}} +b_{i_{1}i_{2} \cdots i_{2n-1}} &\textrm{if
$n=2, 3, \cdots, \frac{N+1}{2}$}
\end{array} \right. , \nonumber\\
F_{i_{1}i_{2} \cdots i_{2n}}(u_{0}) &=& \left\{
\begin{array}{ll}
-3k^2u_{0x}\sum\limits_{W_2}(-1)^{\tau(j_1,j_2)} V_{i_{j_1}}
(V_{i_{j_{2}}})_{u_0} +b_{i_{1}i_{2}} &\textrm{if $n=1$}
\\
3k\sum\limits_{W_3}(-1)^{\tau(j_1,j_2, \cdots ,j_{2n})}
U_{i_{j_1}i_{j_2} \cdots i_{j_{2l}}}
U_{i_{j_{2l+1}}i_{j_{2l+2}} \cdots i_{j_{2n}}} \\
-3k^2u_{0x}\sum\limits_{W_2}(-1)^{\tau(j_1,j_2, \cdots ,j_{2n})}
V_{i_{j_1}i_{j_2} \cdots i_{j_{2l-1}}} (V_{i_{j_{2l}}i_{j_{2l+1}}
\cdots i_{j_{2n}}})_{u_0} +b_{i_{1}i_{2} \cdots i_{2n}} &\textrm{if
$n=2, 3, \cdots, \frac{N}{2}$} \end{array} \right. , \nonumber\\                                                                                                                                                                                     \nonumber\\
\end{eqnarray}
and $$ R(u_0)= \int{\frac{\sqrt{-c_1(c_2k^4+c_1\omega)}}
{(ku_0^2+c_1)\sqrt{2ku_0^3 +\omega u_0^2 +2c_1u_0-c_2k^3}}}\ du_0,
$$ where $u_0$ represents the solution of KdV equation \eqref{Eqn-kdv},
$r_{i_{1}i_{2} \cdots i_{2n-1}}$, $s_{i_{1}i_{2} \cdots i_{2n-1}}$,
$\alpha_{i_{1}i_{2} \cdots i_{2n-1}}$, $h_{i_{1}i_{2} \cdots
i_{2n}}$, $g_{i_{1}i_{2} \cdots i_{2n}}$, $b_{i_{1}i_{2} \cdots
i_{2n-1}}$, $b_{i_{1}i_{2} \cdots i_{2n}}$ are arbitrary constants.

\subsection{Conclusions}

In summary, a simple bosonization approach to deal with
suppersymmetric system is developed. With $n$ fermionic
parameters, the bosonization procedure of the supersymmetric systems
has been successfully applied to the sKdV equation in detail. Such
an integrable nonlinear system is simplified to the usual KdV
equation together with several linear differential equations without
fermionic variables. The traveling wave solutions of the
bosonization systems can be obtained simply by integrations, say,
\eqref{result1} and \eqref{rpl} for the two and three fermionic
parameter cases, respectively.

Some special types of exact
supersymmetric extensions of any solutions of the usual KdV equation
can be obtained straightforwardly through the exact solutions of the
KdV equation and the related symmetries (for instance,
\eqref{result1_1} and \eqref{NSOL2}). The general extensions of the
KdV solutions to those of supersymmetric form are obtained by
solving some linear systems with variable coefficients depending on
the usual exact KdV solutions.

From the procedure exhibited in this paper, we can conclude that the
bosonization approach can be applicable to not only the
supersymmetric integrable systems but also all the models with
fermion fields no matter they are integrable or not. 

It should be emphasized that the solutions obtained via the bosonization procedure
are completely different from those obtained via other methods such as the bilinear approach \cite{BL}.
Especially, the traditional multisoliton solutions of the sKdV are different from ours, say, Eq. \eqref{NSol}.
This fact shows us that for the sKdV equation there exist various kinds of localized excitations. In other words, in additional to the single
supersymmetric traveling wave soliton solution (in the super space-time $\{x,\ t,\ \theta\}$) known in literature \cite{BL},
there are infinitely many single traveling soliton extensions in the usual space-time $\{x,\ t\}$.

The abundant property of the soliton excitations of the classical sKdV reveals some open problems in the both classical and quantum theories.
In the classical level, one of the most important problems may be how to further develop the bosonization procedure such that the known solutions obtained in other approaches can also be included in. In other words, How to developed a bosonization procedure such that the bosonized system is completely equivalent to the classic supersymmetric one. In the quantum level, three of the important topics should be mentioned to investigated: How to reflect the richness of the localized excitations in the usual quantization procedure of the supersymmetry models\cite{Kulish}? Can we find a quantized bosonization approach for supersymmetric integrable systems? What is about the quantization versions of the bosonized systems of this paper?

\subsection{Acknowledgement}

The work was sponsored by the National Natural Science Foundation of
China (Nos. 10735030 and 10905038), the National Basic Research
Programs of China (973 Programs 2007CB814800 and 2005CB422301) and
K. C. Wong Magna Fund in Ningbo
University.

\end{document}